\documentclass[10pt,preprint2]{aastex}

\newcommand{\bgc}{$B_{gc}$}

\shorttitle{The Red-Sequence Cluster Survey}
\shortauthors{Gladders \& Yee}

\begin{document}


\title{The Red-Sequence Cluster Survey I: The Survey and\\
Cluster Catalogs for Patches RCS0926+37 and RCS1327+29}


\author{Michael D. Gladders\altaffilmark{1}} \affil{Carnegie Observatories, 813 Santa Barbara Street, Pasadena, CA 91101\\and\\
Department of Astronomy and Astrophysics, University
of Toronto, 60 St. George St., Toronto, ON, M5S 3H8, Canada}
\and 
\author{H.K.C. Yee\altaffilmark{1}}
\affil{Department of Astronomy and Astrophysics, University
of Toronto, 60 St. George St., Toronto, ON, M5S 3H8, Canada}


\altaffiltext{1}{Visiting Astronomer, Canada-France-Hawaii Telescope,
which is operated by the National Research Council of Canada, le
Centre Nationale de la Recherche Scientifique, and the University of
Hawaii.}


\begin{abstract}
The Red-Sequence Cluster Survey (RCS) is a $\sim$100 square degree,
two-filter imaging survey in the $R_C$ and $z'$ filters, designed
primarily to locate and characterise galaxy clusters to redshifts as
high as $z=1.4$. This paper provides a detailed description of the
survey strategy and execution, including a thorough discussion of the
photometric and astrometric calibration of the survey data. The data
are shown to be calibrated to a typical photometric uncertainty of
0.03-0.05 magnitudes, with total astrometric uncertainties less
than 0.25 arcseconds for most objects. We also provide a detailed
discussion of the adaptation of a previously described cluster search
algorithm (the cluster red-sequence method) to the vagaries of real
survey data, with particular attention to techniques for accounting
for subtle variations in survey depths caused by changes in seeing and
sky brightness and transparency. A first catalog of RCS clusters is
also presented, for the survey patches RCS0926+37 and
RCS1327+29. These catalogs, representing about 10\% of the total
survey and comprising a total of 429 candidate clusters and groups,
contain a total of 67 cluster candidates at a photometric redshift of
$0.9<z<1.4$, down to the chosen significance threshold of
3.29$\sigma$.
\end{abstract}




\section{Introduction}

The detection and characterization of galaxy clusters has long been a
goal of observational cosmology. A large number of surveys over a
broad range of wavelengths have been completed in the past 50 years
\citep[see][for a detailed description of earlier surveys]{bah77}, and
similar searches continue to be done \cite[e.g.,][]{gun86, gio90,
dal92,lum92,sch97,ebe98,ros98,vik98,boh00,bra00,gal00,rom00,ebe01,bah03,gil03,mul03,pie03}. The
goals of modern surveys for galaxy clusters are better defined than
their predecessors, having moved on from the typically cartographic
pursuits of the early days. To be useful in a modern context, a
cluster survey must be well-defined, and must present a homogeneous
and well-understood catalog. These basic requirements come about
because the scientific questions which will be addressed with such
catalogs often require a statistical analysis of large samples
\citep[e.g.,][]{mar01,bah02}. A further criterion for modern cluster
surveys is that they probe a large and distant volume, again motivated
by the type of studies, such as the determination of cosmological
parameters \citep[e.g.,][]{ouk92,lev02}, envisioned with these cluster
catalogs.

It is these joint requirements of a clean, well-characterized sample
and large volumes which led to X-rays being the preferred cluster
search method in the past decade. Because the intensity of
bremsstrahlung emission is proportional to the square of the electron
density, X-rays offer the advantage of selecting only the densest hot
gas found in the deepest potential wells.  Hence, X-ray samples of
clusters tend to be relatively unaffected by projection effects (since
a projection of mass along a line of sight does not emit significantly
at X-ray wavelengths compared to the same mass gathered into a
cluster) and so tend to produce a clean sample. Moreover, it is
possible to survey large areas of sky shallowly with X-ray telescopes
\citep[e.g., the ROSAT all-sky survey,][]{vog99} and thus X-ray
samples have tended to probe larger areas \citep[e.g.,][]{gio90} than
surveys at other wavelengths.  However, though such surveys have
produced extremely useful cluster samples, their mass sensitivity is
ultimately limited by precisely the physical effect which makes them
attractive. It is likely that X-ray surveys will always produce the
largest samples of the most massive clusters \citep{ebe01}, but
unlikely that X-rays will be the most effective approach in probing
extremely deeply into the cluster mass function at redshift one or
higher, or in probing extremely large volumes (i.e., a good fraction
of the observable universe) to high redshifts, where clusters are
generally expected to be less massive, quite apart from cosmological
dimming effects.

As has been suggested by numerous authors
\citep[e.g.,][]{pos96,ols99}, an alternative approach for finding
distant clusters is to use deep optical imaging data. One strategy for
this is demonstrated in \cite{gla00}. \cite{gla00} showed that two
filter imaging is sufficient to perform a clean cluster search using
the cluster red-sequence of early-type galaxies, even when probing
deeply into the mass function. Other techniques exploiting similar
optical data have also been suggested, including the matched-filter
algorithm \citep{pos96} and its variants \citep{kep99,lob00,kim02},
methods relying primarily on searches for the early-type galaxy
population \citep{ost98,got02}, and the search for the unresolved
background light of the cluster \citep{dal96}. Recent application of
some of these methods to even the shallow imaging data of the Sloan
Digital Sky Survey (SDSS) illustrates the potential power and
efficiency of optical cluster surveys \citep{bah03}. A complementary
development in the late 1990s has been the advent of panoramic mosaic
cameras for 4-meter class telescopes; these cameras make it feasible
to image the sky area required to probe a large volume to redshifts
much higher than the SDSS. These two developments, large cameras and
efficient search algorithms, were the impetus for the Red-Sequence
Cluster Survey (RCS).

In this paper we lay out the motivation, design and execution of the
RCS in detail. We pay particular attention to both the photometric and
and astrometric calibration of the RCS images, and measure the
uncertainties in these calibrations both by internal consistency
checks and by comparison to other data.  We provide a detailed
discussion of the adaptation of the cluster finding algorithm of
\cite{gla00} to the complexities of RCS data, with a description of
techniques for accounting for subtle variations in survey depths
caused by changes in seeing and sky brightness and transparency. The
paper concludes by applying this modified algorithm to the calibrated
data from the first two completed RCS patches, RCS0926+37 and
RCS1327+29, which comprise about 10\% of the complete survey. The
resulting cluster catalog is given in its entirety over most of the
RCS redshift range ($0.2<z<1.4$) down to a modest significance cut, 
and includes
richness estimates for each cluster using the $B_{gc}$ statistic
\citep[e.g.,][]{yee99}. Catalogs of clusters for other patches will be
presented in future papers, as will further catalogs for the
patches presented here (less robust catalogs to smaller significance
cuts and refined lower-redshift catalogs using upcoming bluer imaging
data).

This paper is arranged as follows. In \S2 we describe the basic goals
of the RCS, and how the survey was designed to meet these
goals. Section 3 lays out the RCS observational strategy. In \S4 we
provide a detailed description of the data reduction pipeline. We
demonstrate the final data products for two of the survey patches in
\S5. In \S6 we summarize the basis of the RCS cluster finding
algorithm \citep{gla00}, and discuss a number of modifications and
enhancements pertaining to the application of this algorithm to real
RCS data. The cluster catalogs for two survey patches are given in
\S7. We use a $\Omega_M$=0.3, $\Omega_\Lambda$=0.7, and $h$ = H$_0$ /
100 km~s$^{-1}$~Mpc$^{-1}$ cosmology, unless otherwise noted.

\section{Survey Design}

\subsection{Basic Goals}
Five basic goals (finding clusters in a large volume, at high
redshift, and to low masses, with excellent catalog uniformity and
utility) drive much of the survey design for the RCS. The need
for a large volume mandates a survey area of at least 10's of square
degrees, and preferably larger. There are similar sized recent
surveys: a notable comparison, given that it also targets clusters at
$z\sim1$, is the 48 square degree ROSAT Distant Cluster Survey
\citep{ros98}. The local abundance of rich clusters (defined here as
systems corresponding to Abell Richness Class 1 or greater, or
equivalently those systems with a one-dimensional velocity dispersion
in excess of 750 km~sec$^{-1}$) is on order of one every $10^5-10^6$
$h^{-3}$~Mpc$^3$ \citep[e.g.,][]{bra00}. Over the interval
$0.5<z<1.0$, the total comoving volume per square degree is about
0.5-1.0$\times$10$^6$ $h^{-3}$ Mpc$^3$, depending on cosmology, and so
one expects on order of one such cluster per square degree in this
redshift interval, and likely less since the cluster mass function is
expected to be reduced at higher redshift. Given these issues, and
considerations of feasibility given limited telescope and researcher
resources, we initially chose the RCS size as 50 square degrees, all
to be imaged using the Canada-France-Hawaii Telescope (CFHT). With the
addition of several new collaborators, we added another 50 square
degrees to be imaged at the Cerro-Tololo Interamerican Observatory
(CTIO) 4m telescope, bringing the total planned survey area to 100 square
degrees.

When the RCS was initially designed, there were very few clusters
known at $z>0.7$. Most examples at that time, apart from a few very
massive clusters from the EMSS \citep{gio94}, came from the survey of
\cite{gun86}. Hence, from the perspective of survey design, it was
clearly important to attempt to find a significant cluster sample at
$z>0.7$. Moreover, the samples at $z>1$ were extremely small - only a
handful of clusters - making this redshift regime particularly
significant. Most importantly, the highest redshift clusters have the
most significance for determining cosmological parameters
\citep[e.g.,][]{lev02}, and offer the best constraints for studies of
the evolution of cluster galaxies.  We chose redshift one as a
fiducial target redshift for these reasons.
%Also, note that when performing an
%imaging survey with any telescope there is a natural ``sweet spot''
%for efficient imaging which balances inefficiencies due to
%observational overheads against the inefficiency of imaging too
%deeply\footnote{I.e., once the observations become sky noise
%dominated, it requires 10$\times$ the integration time to probe 1.25
%magnitudes fainter.}.  Typically the most efficient integration times
%are on order of $\sim10$ minutes. 
The bulk of the signal for cluster
finding comes from cluster galaxies within a couple of magnitudes of
M$^*$ \citep{gla00}; at redshift one this depth can be reached efficiently
using 4m class telescopes. This depth is also needed to be sensitive
to lower mass clusters at more moderate redshifts \citep{gla00}.

The high redshift goal of the RCS also mandates using very red
filters (recall that the cluster finding algorithm of \cite{gla00} is
optimal when the filters span the 4000\AA~break) and we hence chose
to use the $R_c$ (centered at $\sim$6500\AA) and $z'$ (centered at
$\sim$9100\AA) filters. The color magnitude diagrams in Figure 1 show
why this choice is important.  Plotted are fiducial red-sequences in
AB magnitudes for clusters from $0.5<z<1.4$ using the $V$ and $I$
filter pair often utilized in optical cluster surveys
\citep[e.g.,][]{pos96} and using the $R_C$ and $z'$ filter pair
adopted for the RCS. All filter curves used are for the CFH12K camera
\citep{cui00}, with normal rather than red-sensitive CCDs (see \S2.2
below). The model used is a GISSEL \cite{bru93} model parameterised as
a 0.1 Gyr burst ending at $z=2.5$ with a $\tau=0.1$ Gyr exponential
decline thereafter, in an $h$=0.7 universe. The $R_C$ and $z'$ filter
pair clearly provides much better color discrimination at high
redshift, as the colors are non-degenerate all the way to $z=1.4$, and
more widely separated at $0.5<z<1.0$. Moreover, the $z'$ and $R_C$
filters suffer less drastic K-correction dimming until redshifts
well above one, with typical early-type galaxies at $z$=1.4 being 0.9
and 1.2 AB magnitudes brighter than $I$ and $V$, respectively. Even
the fact that the $z'$ filter is generally less efficient than $I$
(due to falling CCD response in the red and a brighter sky) is also of
no import, since the limiting filter at $z>1$ for either filter set is
the bluer one.

The basic outline of the RCS is thus a 100 square degree survey in
$R_C$ and $z'$ to a depth $\sim$2 magnitudes past M$^*$ at redshift
one. This design provides a cluster sample over a large volume, to
high redshifts, and low masses. By completing the survey on only two
telescopes, using a single imaging instrument at each, we also hoped
to ensure uniformity in the data products. Finally, to enhance the
utility of the survey, we divided the survey area into a number of
individual patches, which are placed to allow maximum flexibility in
observing and follow-up. In practice, we chose ten patches for the
CFHT component of the survey, and twelve patches for the CTIO
component. The rest of this paper focuses primarily on a subset of the
CFHT observations - namely, the first two completed patches. The
remainder of the CFHT data and the CTIO data are discussed further
elsewhere \citep[e.g.,][]{bar02}.

\subsection{Patch Layout}

The instrument used for the CFHT observations is the CFH12K
\citep{cui00}. This camera is a mosaic of twelve 2k$\times$4k CCDs
arranged in a 6$\times$2 grid, with typical inter-chip gaps of seven
arcseconds.  The camera has a plate scale of 0.206 arcseconds per
pixel, corresponding to a $42\times28$ arcminute\footnote{Throughout
this paper, when sizes of fields are described, the size in RA is
given first, followed by the size in DEC.} image for the full
mosaic. All chips in the CFH12K are from the MIT/Lincoln Labs CCD
project. The CCDs come in two varieties: ``standard'' chips, and
``red-sensitive'' chips which are thick, high-resistivity devices with
enhanced response at wavelengths redward of $\sim$7000\AA.

The patch size for the CFHT observations was chosen to provide ten
equal patches, with each consisting of fifteen CFH12K pointings
arranged in a slightly overlapping grid of $3\times5$ pointings. With
a 30 arcsecond overlap this corresponds to a patch of size
$125\times138$ arcminutes. These patches were placed according to a
number of considerations. The first was that we wished to overlap some
of the patches with regions covered by other galaxy/galaxy cluster
surveys. This provides added value either through comparison to
cluster search results via other methods, or via complementary
data. Secondly, we wished to avoid regions of significant interstellar
dust, as the resulting extinction degrades both the depth and
uniformity of the survey. We also wished to avoid bright stars, as
these result in lost area, and make the data processing more
difficult. These two considerations set a general constraint that all
the patches be located at galactic latitudes greater than
40$^\circ$. Finally, we also avoided too high a galactic latitude to
ensure that there are enough reference stars for star-galaxy separation,
astrometric corrections, and detailed point-spread-function
corrections for lensing analyses \citep[e.g.,][]{hoe02}.

The precise location of each patch was chosen by a careful comparison
to the integrated HI map of \cite{har97} and the star counts from the
Tycho catalog \citep{hoe97}.  The HI column densities were converted
to an extinction estimate using the conversion from \cite{bur78}.  To
place a patch, three maps spanning the relevant portion of the sky
were produced with a 0.1 degree grid spacing. The first map is an
average of the estimated $E(B-V)$, convolved with a kernel the same
size as a fiducial patch. The second map similarly records the
integrated flux of all stars down to $R_C=9$. The $R_C$ magnitudes
were estimated from the $B$ and $V$ magnitudes reported in the Tycho
catalog. The limit of $R_C=9$ is set both by the completeness depth of
the Tycho catalog, and by the fact that the density of stars at
$R_C=9$ is high enough that the total light from fainter objects is a
smooth function of galactic latitude for the regions
considered when smoothed on the patch scale. The third map simply
records the brightest star in the patch area in the $R_C$ band.
Using these maps, we then searched for locations which had low average
$E(B-V)$, preferably less than 0.05, no star brighter than $R_C$=6 and
with the brightest star as faint as possible, and low total light from
bright stars.  Typically, there were a number of candidate placements
for each patch; as a final step each was examined visually for bright
galaxies (which were avoided) and then the placement which best
matched the criteria outlined above was selected. For patches which
were to be placed overlapping areas from pre-existing surveys, we
constructed similar maps over the much smaller allowed area with a
finer grid, and fine-tuned the patch placement to maximize the overlap
while minimizing the extinction and avoiding bright stars.

Figure 2 shows an example of the patch placement, for the patch
RCS1327+29.  The background image is the digitized Palomar Sky Survey
image for the region, and the over-plotted lines show the placement of
the 15 CFHT pointings. The pointings within each patch are designated
by a row and column code; the columns run from A to C, with A to the
east, and the rows run from 1 to 5, with 1 to the north. This
convention holds for the entire survey, with modifications in cases
for extra or missing data. Note also that the 1327+29 patch was chosen
to overlap a much smaller patch from the Palomar Distant Cluster
Survey \citep{pos96} and a patch from the older GHO survey
\citep{gun86}. In all cases such as this where we overlapped areas
with known surveys, we ensured that the original patch definition in
the older surveys was random with respect to clusters, and we did not
use the location of known clusters in the area to guide the patch
placement. This ensures that the resulting patch is unbiased with
respect to galaxy clusters.

The central coordinates for all ten CFHT patches are listed in Table
1.  In each case we tabulate the 100$\mu$m brightness estimated from
IRAS maps, and the estimated average extinction from \cite{sch98}.

Several of the observed patches deviate from the nominal plan of
fifteen full pointings of the CFH12K camera. Patches 0926+37, 1327+29,
1415+53, 1615+30, and 2151--06 all were observed in the first run,
during which the CFH12K camera was missing two chips. Hence these
patches are missing some area. Due to scheduling requirements, we were
also unable to complete two pointings in patch 1447+09 and three in
2151--06, though we did acquire three extra pointings in patch
0920+37. As a result, the entire CFHT component of the survey covers
$\sim$46 square degrees. For each patch, the area in square degrees is
indicated in Table~~1.

\section{Observational Strategy}

The RCS observational strategy is notably different from most ongoing
surveys using mosaic cameras, and hence worth describing. The most
obvious difference is that the observations are not dithered: for the
data from CFHT a single 15 minute $R_C$ integration was taken at
each position, as well as two 10 minute $z'$ exposures without
shifts. The 20 minutes of $z'$ integration was split solely to keep
sky levels at a reasonable value.  This minimalist approach, which
does not allow for the rejection of cosmetic defects or cosmic rays in
the images, is driven by the need for observing efficiency and
simplicity in the data processing. This latter point is of particular
note, and is explored further in \S4.  The presence of cosmic rays and
defects in the images results in a minimal loss of area, a loss which
is more than compensated by the high efficiency allowed by this
observing mode. Cosmic rays in particular can also affect the
photometry of a small number of objects. However, the cluster finding
algorithm used on these data is insensitive to these effects.

Typically, each pointing was observed all at once, with the three
integrations taken sequentially. The only exception to this is
pointings imaged near twilight. Experience from the first CFHT run in
May 1999 showed that $z'$ images taken near twilight do not defringe
as well as those taken in the middle of the night. Hence on all subsequent
runs, we typically observed two pointings in $R_C$ sequentially at the beginning
and the end of the night, with the corresponding $z'$ data acquired
more towards midnight. In most cases, pointing was done blindly, since
slewing to a target directly and integrating without checking the
pointing is the most time efficient. For pointings with split $R_C$ and
$z'$ data we ensured that the data were taken at the same position by
returning to the same telescope coordinates and guider position. This
ensured a good position match in all but a few cases.

Apart from these two changes, the observing runs proceeded in a fairly
standard manner. Photometric standard fields from \cite{lan92}
were observed during twilight at the beginning and end of each
photometric night, and the central region of M67 was observed once per
run for astrometric calibration. In cases where data were deemed
non-photometric we acquired short integrations of the same field
 during a photometric night to ensure a proper calibration.  The basic
data for each run are summarized in Table~~2. The RCS runs were
mostly photometric, typically with sub-arcsecond seeing. One complete
night was lost during Run 3, and the equivalent of approximately one
more night was lost during the remaining runs due to telescope
problems, and minor weather losses. In total, the entire imaging
program required 11 clear nights.

The data discussed in the rest of this paper are from Run1-a, on May
5-6 1999, Run 2, on January 7-14, 2000, and Run 4, on January 27-28,
2001. A total of 33 pointings comprising two patches are considered,
and the relevant data for each, including estimates of the seeing for
each pointing, are given in Tables 3 and 4.

\section{Data Reduction}
The data reduction for mosaic images is typically quite complex, and
fraught with a number of subtleties which can hinder those used to
working on single-chip CCD data. The design of the RCS observations allows
us to circumvent many of these issues, since each camera chip may be
treated independently. Hence, standard single CCD methods (and
programs) may be used for much of the data reduction. Note that it is
never our goal to construct large scale homogeneous images from the RCS
survey data, and hence photometric and astrometric calibration can be
performed after the data have been extracted to catalog form. This
avoids many of the complications typical of mosaic data, which
often relate to how to photometrically and astrometrically map
different portions of the images into a standard frame in order to
stitch together dithered observations.

The transformation of the RCS survey data from raw images to final
photometrically and astrometrically calibrated catalogs consists of
three major steps. The first is pre-processing, in which typical
procedures such as de-biasing and flat-fielding are performed. The
second major step is object-finding and photometry. The final step,
performed using the individual chip-by-chip catalogs output from step
two, is to stitch the individual catalogs into a master catalog for
each patch using a full photometric and astrometric calibration.  Each
step consists of a large pipeline, written specifically for these
data.

\subsection{Pipeline I: Pre-Processing}

The pre-processing of the RCS survey data was done, for the most part,
in the standard manner, using a pipeline implemented within
IRAF\footnote{IRAF is distributed by the National Optical Astronomy
Observatories, which are operated by the Association of Universities
for Research in Astronomy, Inc., under cooperative agreement with the
National Science Foundation.}. Each night, or at least during each
run, we acquired sets of bias, dark and twilight flat-field images.
Both the bias and dark frames contain very little signal, and so we
examined several possibilities in removing their effects. After some
experimentation it was found that removal of the dark frame did
nothing to improve the uniformity of the images. Moreover, some of the
chips contain some columns with significant structure which was best
removed using only bias subtraction (and was often degraded if the
dark was also subtracted), and so we settled on simply removing the
bias and making no dark current corrections to the images. However, we
did continue to acquire dark images in later runs, in order to monitor
possible changes in the dark current.
 
\subsubsection{$R_C$-band Images}

The $R_C$-band images were processed simply by overscan subtraction,
de-biasing, and flat-fielding using twilight flats, all in the
standard manner.  As a final step for data from each night, all
available $R_C$-band images were combined using rejection algorithms to
produce a super-flat. This was smoothed to eliminate small scale
noise, and all relevant images were re-flattened using this
super-flat. The resulting $R_C$-band images typically have variations in the sky
of less than 0.3\% over a single chip.
 
\subsubsection{$z'$-band Images}

The $z'$-band images were overscan corrected, de-biased, and
flat-fielded using twilight flats similarly to the $R_C$-band images.
However, the $z'$ images suffer from significant fringing effects, so
extensive further processing was required. 
%Fringing in CCDs is an
%additive background signal which results from multiple scattering
%within the CCD itself, and is most visible when the CCD is strongly
%illuminated by monochromatic light (such as the strong night sky lines
%red-ward of 7000\AA) rather than a smooth continuum source. 
For the CCDs
used in the CFH12K, the fringe amplitude can exceed 10\% of the total
sky signal on the worst CCDs. 
%To subtract the fringe signal, it is
%necessary to construct a fringe frame, which is akin to a
%superflat. This frame is taken as representative of the fringe
%pattern, and is subtracted from the target image after scaling to
%match the fringe amplitude on the image in question. In practice, it
%is typical to first fit and divide the fringe frame by a low-order
%surface polynomial; this fit is considered a superflat and is also used
%to re-flat-field the target image prior to defringing.

One additional complication with de-fringing these data is that the
fringe pattern is not completely stable either in time or across the
sky, since natural spatial and temporal variations in the night sky
lines cause variations in the corresponding fringe patterns. Apart
from variations in the fringes, this can be seen in large variations
in the $z'$ sky brightness. Even in the absence of moonlight, the $z'$ sky
brightness was observed to vary by factors of 2-3 over a single
night. Often this means that the data from an entire night cannot be
combined to produce a fringe frame, unlike the case when producing a
superflat. After much experimentation, we settled on a procedure for
producing fringe frames in which the fringe frame for a given $z'$
image was constructed from a weighted sum of the all $z'$ frames from
that night. The weighting is done according to time, location in the
sky, and the overall sky brightness. Frames which are closest in time
are given the most weight, typically using a Gaussian weighting
function with a sigma of $\sim$3 hours. Images from the same patch,
which are hence nearby, are also given twice the weight of other
frames.  Finally, input images were also weighted by the difference in
the sky levels; for example an image of half or twice the sky
brightness of the target image is assigned half the weight of an image
of the same sky brightness. 

The standard image combining algorithms in IRAF were found to be
insufficient for constructing the weighted fringe frames, and so
further code was written to implement a two-step image combine. In
this process, all the images to be used in a fringe frame are first
approximately flattened using an unweighted fringe frame estimated
from the entire night. All pixels brighter than three times the
standard deviation in the sky pixel values were then masked in each
frame, with the masking significantly padded to exclude the low-level
extended halos of bright objects. Masked versions of the original $z'$
images were then combined using appropriate weights to construct the
fringe frame for each observation.

Each resulting fringe frame was then smoothed with a 3$\times$3
boxcar, and then the appropriate scaling with which to subtract the
fringe frame was determined by an iterative analysis of the sky
residuals for different scalings.  The masked regions established
above were also used in this step to ensure that only sky pixels were
used in this process.

The resulting de-fringed $z'$ images are generally very flat, with
residual fringes typically having an amplitude of less than 0.5\% of
the sky. On occasion the fringe residuals are higher - on order of
1\%. Further experimentation using different weighting schemes and
fringe removal strategies did not result in significant improvement in
these frames.

\subsubsection{Corrections for Saturation Effects}

One final unusual step in the pre-processing of the CFH12K data is a
correction implemented for saturation effects. Proper recognition of
saturated pixels is important in the object-finding and photometry
process, as these pixels must be excluded. 

The CFH12K has a number of CCDs which saturate rather strangely. On
these CCDs the count value on a saturated pixel actually drops as
more charge is gathered, i.e., so that the most saturated portion of a
saturated object is in fact measured at a lower count value than less
saturated regions.  Furthermore, once an object saturates to the point
of bleeding, the bleeding columns produced are not at the saturation
level of the CCD but at a much lesser level which in some cases is
below the sky level. This makes it non-trivial to establish what
pixels are affected by saturation using typical analysis programs.

To circumvent this problem, a separate masking algorithm was
established as part of the IRAF pre-processing pipeline. This
algorithm is designed to detect and isolate saturated objects by
keying on those pixels which are just saturated and hence recognizable
because they fall above some pre-determined threshold below the actual
saturation level. Using these starting pixels, each saturated object
was then traced out to a much lower threshold which was empirically
determined to be lower than the typical bleeding level for the chip in
question. All pixels above this lower threshold are deemed to be
saturated, and each object was then reconstructed so that the most
central saturated pixels for each object had the highest value, and a
value well above the saturation point for the chip. This produces
images in which saturated objects appear similar to those on typical
CCDs, with the precise manner in which the objects are reconstructed
tailored to ensure that photometry pipeline works smoothly. Neither
the photometric nor astrometric information for these
``reconstructed'' objects is correct in detail, however, and we are
careful to exclude them from any further analysis. The reconstruction
process serves simply to streamline a number of later elements of the
reduction pipeline.

\subsection{Pipeline II: Object-Finding and Photometry}

Object-finding and photometry were performed using the package PPP
\citep{yee91} modified to operate in a pipeline mode.
\cite{yee91} provides detailed descriptions of the algorithms used
along with analysis of simulated data demonstrating the characteristics
of the algorithms.
Further improvements to PPP, motivated by the need for photometric 
reduction of large imaging database, are described in
in \cite{yee96}, as well as \cite{yee00}. 
We refer the reader to these papers for more details, and present
here only a very brief overview of the methods used.

\subsubsection{Object Finding}
Object finding was performed on weighted sums of the $R_C$ and $z'$
images with the weights based on the signal-to-noise ratio of the
input pixels.  The $R_C$ and $z'$ images were registered prior to
summation by using bright but unsaturated objects in each image. Due
to the data acquisition strategy, shifts were typically only a few
pixels, for which a simple offset is sufficient.  This stacked image
was then masked to exclude known hot columns and cosmetic defects, and
the bleeding columns from saturated stars.  Diffraction spikes are
also mapped and excluded based on the positions of bright saturated
stars.  The image is then smoothed using a 3$\times$3 tapered
smoothing box, and all peaks with a net flux averaged over a
3$\times$3 box greater than 2.6$\sigma$ of the smoothed local sky are
selected. This limit corresponds roughly to a 1$\sigma$ threshold in
the unsmoothed image. Each detection is then subject to a number of
tests, including a ``sharpness'' test on the unsmoothed image to
reject cosmic rays with full-width-at-half-maximum smaller than two
pixels, and a size test to reject objects resulting from noise spikes
smaller than a point source.  The resulting object list, typically
4000-6000 objects per chip, was then eye-checked to ensure that the
various masking procedures for diffraction spikes, bleeding columns
and other cosmetic defects have performed properly.

\subsubsection{Photometry}

The photometry pipeline produces a total magnitude estimate for each
detected object in the deeper of the two filters (usually the $R_C$
filter, except for very red objects) by analysing its photometric
curve of growth, which is constructed from a pre-defined set of
circular apertures, with masking of nearby objects as required.  In
all cases the total magnitude is measured at an ``optimal aperture'',
deduced by analysing the shape of the curve of growth
\citep[see][]{yee91}.  This magnitude within the optimal aperture is
then corrected to a standard aperture of 8.5 arcsecond diameter
(functionally indistinguishable from an infinite aperture for a point
source) for objects with the optimal aperture smaller than the
standard aperture, using corrections derived from the shape of the
growth curves of bright point-source objects.  For bright galaxies of
larger size, a growth curve up to a maximum diameter of 25 arcseconds
is used to determine the optimal aperture to make sure that the bulk
of the light is included.  We note that very bright galaxies
($R_C<\sim15$ mag) will in general have underestimated magnitudes due
to their larger sizes.

The color of each object is estimated separately from the
total magnitude, in an aperture (called the color aperture) of either
three arcseconds, or the optimal aperture, whichever is smaller. The
total magnitude for the second filter is then computed from the color
and the total magnitude of the first filter \citep[see][]{yee91}.
This assumes that the color gradient within each
galaxy produces a negligible effect in the relative total magnitude.

The final photometric catalog for each chip of each pointing has
errors on the magnitudes derived from the photon noise in the optimal
aperture for each object, summed in quadrature with an ``aperture
error'' of 0.03 magnitudes. This extra error accounts for the uncertainty
in estimating the optimal aperture \citep{yee96}. Because the galaxies
of interest are sky-noise limited, the photometric uncertainty is
computed based on the sky noise within the aperture. The color error is
the sum in quadrature of the photon noise for each filter in the
color aperture. Using a relatively small color aperture minimizes
the error in the color.

\subsubsection{Star-Galaxy Separation}
Star-galaxy separation is performed by comparing each object to a
local set of bright but unsaturated reference stars
\citep{yee91,yee96}. The star-galaxy separation for a typical field,
showing the object compactness versus magnitude, is shown in Figure
3. In all cases the star-galaxy separation is robust for all but the
faintest objects (those below the $\sim100$\% completeness limit),
once the process is checked by eye to eliminate occasional problems
with the automatic reference star selection. Star-galaxy separation is
performed in both filters separately. The $R_C$ filter is used for the
primary classification, since it is generally deeper, but objects
which have significantly higher S/N in the $z'$ image are classified
using the $z'$ image instead. Note that any cosmic rays that have
passed through the initial object-finding are eliminated by the
star-galaxy separation at this stage as they are typically smaller
than the measured point spread function.  In the final photometric
catalog, objects are classified into four categories: galaxies, stars,
saturated stars, and spurious ``non-objects'' (e.g., cosmic ray
detections, cosmetic defects, etc.).

\subsection{Pipeline III: Master Catalog Assembly and Calibration}

The final step in the RCS data processing is to assemble the
chip-by-chip catalogs (in instrumental magnitudes and pixel
positions) into an astrometrically and photometrically calibrated
master catalog for each patch. This process has a number of steps,
detailed below.

\subsubsection{Photometric Calibration}

%The photometric calibration of
%CCD images is essentially a process of relating CCD counts to a
%physically meaningful and well calibrated flux scale. This is
%typically accomplished by observing a number of so-called ``standard''
%fields in the course of an observing run, and using objects of known
%brightness in these fields to establish the required
%transformations. Procedures for this are well established. 

The photometric calibration of the RCS data is complicated by the
mosaic cameras and variations in those cameras from run to run, and
hence a matter for significant discussion. Uniform photometric
calibration is particularly important for the RCS, since the accuracy
of photometric redshifts is limited by the systematic uncertainty in
object colors. Absolute calibration to a particular system is of less
concern; uniformity in the photometric calibration ensures accurate
photometric redshifts so long as clusters of known redshift are
observed within the survey area.  Given the scope of the project we
have not attempted to measure higher order calibration terms (such as
color-dependent airmass terms), since the required effort in gathering
standard data would overwhelm the actual gathering of science data in
such a case.

With observations in two filters (call these filters $i$ and $j$), and
at the level of precision achievable using the standard data available
within the RCS, the goal is to solve the following equation for each
object:

\begin{eqnarray}
m_i=m_{Ii}+A_{0i}+A_{1i}\times (Airmass)+\nonumber\\ A_{2ji} \times (m_j-m_i)
\end{eqnarray}
\noindent where
\begin{quote}
$m_i$, $m_j$ are the magnitudes of the object in the standard system in 
the filters $i$ and $j$

$m_{Ii}$ is the instrumental magnitude of the object in filter $i$

$A_{0i}$ is the zero-point in filter $i$

$A_{1i}$ is the extinction coefficient in filter $i$, and

$A_{2ji}$ is the color term transformation coefficient in filter $i$,
referenced to the $j-i$ color. \end{quote} A similar equation governs
filter $j$. 
%To compute the desired quantities for unknown objects
% ($m_i$ and $m_j$) one must establish the values of $A_{0i,j}$,
% $A_{1i,j}$ and $A_{2ji,ij}$. This is the primary goal of the
%photometric calibration. 

The calibration of CCD mosaic data has the complication
that the chips of the mosaic are essentially independent cameras for
the purposes of calibration, and so the 3 basic coefficients needed, per
filter, become 36 coefficients for the whole CFH12K mosaic, per
filter.  Since standard fields are not arranged with mosaics in mind,
it is then non-trivial to acquire enough standard data to measure all
these coefficients. Fortunately, we expect some of the coefficients to
be very similar. First, the airmass terms should be essentially the
same for all chips. Secondly, provided that the individual CCDs have
similarly shaped quantum efficiency (QE) curves, the color terms
should all be similar. For the CFH12K mosaic, which has two types of
CCDs with markedly different QE curves, we can thus reduce the problem
to establishing a single airmass term, two color terms, and twelve
zero-points - per filter. 

Observations with the $z'$ filter suffer from a further complication,
namely that the filter has until recently been little used, and so the
standards available in the literature are rather limited. Since the
beginning of RCS observations in 1999 this situation has improved
significantly, thanks wholly to the efforts of the SDSS in
establishing a new network of primary and secondary standard fields \citep{smi02}
which include the $z'$ filter \citep{fuk96}. The strategy at the
telescope for acquiring standards hence evolved from run to run,
and moreover, the CFH12K was reconfigured with some new chips between
Runs 1 and 2. We note the individual details for each run below.

\subsubsection{Photometric Calibration: Run1}

At the time of the first run in May 1999 the SDSS standard
star fields were not known to us, and so we relied upon older standards
from the literature. The $R_C$ filter was calibrated by observations
of \cite{lan92} fields. Fields SA101 and SA110 were repeatedly
observed during the run.  The $z'$ standards BD+17$^\circ$4708, Feige
34, Ross 484 and Ross 711 (D. Schneider, private comm.) were also
observed over the course of the run. In this case the star
BD+17$^\circ$4708 was tiled across the entire mosaic by repeated
observations offset to each chip, and other standards were
observed when possible.  The standard fields were pre-processed using
the same pipeline as the science frames, save that the superflat used
was the one deduced from the science observations rather than from the
images of the standard fields. Standard stars were located and their
brightness in a large aperture measured by hand using tools within
IRAF, with aperture corrections deduced from bright isolated objects
applied to the magnitudes of fainter or somewhat crowded objects for
which large measurement apertures were inadvisable. 

Despite the allocation of significant observing time at the telescope,
particularly to tile the mosaic with $z'$ standards, the total number
of standard star measurements was found to be insufficient to measure
all the required coefficients with good accuracy. In most cases
zero-points could be measured individually, but the data for each chip
were insufficient to measure either a color or airmass term. We thus
adopted a compromise strategy in which we rescaled each chip to the
same sky value (using the mean offset in sky values between the chips
established from all of the science frames), allowing us to look
at the observed standards in ensemble. Notably, this rescaling by the
sky value gives chip-to-chip offsets which are within a few percent of
the offsets implied by the zero-points, hence validating the implicit
assumption that the sky has a color similar to the standard
stars. Formally, the residual standard deviation in the zero points
between chips (after re-scaling by the sky) is 0.04 magnitudes in
$z'$, and 0.02 in $R_C$. Notably, the two chips for which multiple
$z'$ standards were observed showed internal standard deviations of
0.04 and 0.03 magnitudes between standards, from which we conclude
that the residuals in the standard measurements after re-scaling are a
good estimate of the uncertainty in the calibrations. The final
zero-points for both filters are thus taken as rescaled versions of
the global zero-point deduced from all chips. Also, note that the $z'$
data have been calibrated to the SDSS system, since the $z'$ standards
observed are also part of the SDSS primary standard set.

Further analysis of this ensemble distribution of standard star
measurements demonstrated that the airmass range explored was
insufficient to measure the airmass term with good accuracy. The
airmass terms used thus come from prior knowledge of the Mauna Kea
site, and in $R_C$ the airmass term is the same as that used for the
CNOC2 survey data \citep{yee00}. To estimate the $z'$ term, we used the
scaling between $r'$ and $z'$ from \cite{fuk96}, applied to the $R_C$
term. Further accuracy is not required, since the highest airmasses in
the science observations are approximately 1.5.

The available standard measurements were also used to estimate color
terms for the two chip types. Over a range of several magnitudes in
$R_C$-$I_C$ color for the Landolt standards, there was no measurable
color term in $R_C$ for either chip type. The $z'$ standards used span
a range of only $\sim1$ magnitude in $i'-z'$ and we similarly could find no
color term from these data. Much more extensive data from Run 2 refine
these conclusions: the color term in $R_C$ is maximally about 0.001
and consistent with zero, and there are small color terms in $z'$
which differ between the chip types. As we expect the color terms to
be stable from run to run, we have retroactively applied the color
terms in $z'$ from Run 2 to the data from the appropriate chips in Run 1. 

\subsubsection{Photometric Calibration: Run 2 and Run 4}

The photometric calibration data acquired during these two runs are
more extensive than the data in Run 1, both because the run was nearly
four times as long, and also because the preliminary SDSS calibration
lists were available to us by that point. For these runs, only Landolt
fields which included SDSS primary standards were observed. Typically,
each field was observed more than once during each run, and several
fields were observed at several camera positions in an attempt to
ensure that some of the relatively sparse SDSS primary standards fell
onto each chip.

These data were pre-processed in a manner similar to the data in Run
1. However, noting that the volume of standard data was growing
rapidly, we implemented a new procedure for measuring the standards.
Essentially, all the standard fields were run through the same
object-finding, photometry, and astrometric calibration procedure
(described below) as the science data. This ensures that these data
are measured uniformly, and also allows us to use automatic algorithms
to match the measured objects to the standard lists.

The resulting $R_C$ calibration data for Run 2 (Run 4 is similar) are
shown in Figure 4. The vertical axis shows the offset between the
computed magnitude (after calibration) and standard magnitudes, versus
the $R_C$-$I_C$ color from the Landolt catalog. The ensemble for
``normal'' and ``red-sensitive'' chips are shown, demonstrating the
lack of significant color terms in both cases. The error bars shown
are the sum in quadrature of the errors in the Landolt catalog and the
measurement errors in the RCS data, and hence there may be additional
scatter due to errors in the calibrations on each chip. However, for
most chips there are a number of observations of standards, allowing a
good estimate of the mean offset. The uncertainty in the mean
calibration for each chip is typically in the range 0.015-0.030
mags. These values indicate the fundamental accuracy on a chip-by-chip
basis.

Despite a greater number of primary $z'$ standards, and the large
number of standard field observations, we found that the $z'$ data
from Runs 2 and 4 were, like Run 1, also insufficient to fully
characterize the $z'$ photometric solution for the CFH12K.
%We expect this problem
%has been fixed in later runs, in which we implemented a rigorous
%tiling strategy to ensure uniform coverage of the twelve chips. 
As in Run 1, we use somewhat non-standard methods to get around these
difficulties. Essentially, we rely on the SDSS data to provide the
$z'$ calibration, using data not from the primary standards list, but
rather from early release SDSS photometry \citep{sto02} of the entire
Landolt fields.  Despite the fact that these data are not of standard
stars per se (in that they have not been measured repeatedly under
well controlled circumstances), the overall excellent quality of the
extensive SDSS photometric calibration program ensures that the data
are perfectly usable in this manner provided that enough objects are
included to guard against secular variations and the like. For
example, by a direct comparison of all stellar objects in overlapping
calibration fields in Run 2, we derive the calibrations shown in
Figure 5. Clearly, the calibration is very good. The zero point for
each chip can be established with excellent precision. A examination
of the zeropoint derived from individual observations of standard
fields indicates that the ensemble zeropoints are accurate to better
than 0.01 magnitudes. However, the RMS uncertainty from observation to
observation is typically 0.02 magnitudes. This latter value is
indicative of the calibration uncertainty in $z'$ for any given
science field observed even under photometric conditions, due likely
to variations in atmospheric transmission on relatively short
timescales. Note also that there is a small color term in $z'$, which
can be measured individually for each chip. Since we expect the
primary difference in color terms between chips to be between the two
chip types, we have in practice measured two color terms (one for the
nine normal chips and one for the three red-sensitive chips).

As a final check of the $R_C$ calibrations we have completed the same
comparison to the SDSS early release data in the SDSS $r'$ filter. As
expected there is a significant color term, since the $r'$ filter is
bluer than $R_C$. There are also uncertainties in the matching of the
zero-points at the level of a few hundredths of a magnitude.  This is
well within the expected uncertainty of the initial calibration to the
Landolt fields. We have thus adjusted the $R_C$ zero-point
calibrations (but not color terms) to minimize the scatter with
respect to average zero-point offset between $r'$ and $R_C$, since the
zero-point in $r'$ is better determined. Finally, note that the
measured color terms when calibrating to $r'$ are essentially the same
for both chip types (unlike the $z'$ filter), implying that, as
expected, the differences between the chip types are insignificant at
these wavelengths.

Finally, we note that the SDSS data used to calibrate these runs are
not the final SDSS photometric data (D. Tucker, private
comm.). Formally, these data are on the $u^*g^*r^*i^*z^*$ system. This
system differs from the planned final SDSS system in that color terms
have not been applied to the data (due to unresolved issues in
transforming the SDSS monitor telescope and primary telescope cameras
to the same photometric system). Systematic differences between the
two systems are expected at the level of 0.05
magnitudes. Re-calibration of these data will thus be required at some
future date, once the SDSS system is fully defined. By then, we expect
several other calibration possibilities to exist - either calibration
to the SDSS system by direct comparison of the science frames, or by
comparison of the standard fields observed by the RCS to the matching
SDSS Secondary Calibration Fields. We leave these possibilities to a
future paper, coincident with the planned release of the RCS
photometry and processed imaging. 

The final calibration uncertainties in the photometric calibration for
Runs 2 and 4 are about 0.02 magnitudes for the zeropoints in each
filter, and 0.001 in the color terms. The color of any given object
may thus have systematic uncertainties of about 0.03 magnitudes in
$R_C-z'$ color. Run 1 has a somewhat higher uncertainty in the $z'$
calibration of 0.04, and correspondingly higher uncertainty in the
color. The mapping of color onto redshift shown in Figure 1 shows that
in either case this corresponds to a systematic redshift uncertainty
of about 0.01-0.015 for moderate redshifts, and up to 0.05-0.08 at
higher redshifts where smaller color changes correspond to larger
redshift differences. For purposes of finding clusters, this
systematic uncertainty is in all cases comparable to the random
uncertainty due to noise in the photometry and uncertainty in the
cluster red-sequence modeling \citep{gla02a}.

\subsubsection{Astrometric Calibration and Catalog Assembly}

The RCS data are astrometrically calibrated using a two-step
process. First, images of M67 taken during the run are used to
establish the placement of the chips relative to one another within
the camera. This is done by running these images through the pipelines
described above, and then matching the resulting catalogs to
astrometry from the USNO-A2.0 catalog \citep{mon98}. Each chip is mapped
into the camera reference frame separately using a second order surface
polynomial with cross-terms. Typically 100-200 stars are used on each
chip to establish this mapping. This first step stitches all twelve
chips of the camera into a common reference frame, and naturally
incorporates such effects as camera distortions and rotation. Since the
camera has random rotation offsets of about a degree from run to run,
this last point is particularly relevant. We also use this distortion
map to establish the variation in pixel size across the camera.

The variation in pixel size across the mosaic causes a
position-dependent bias in the photometry which must be fixed at this
stage. This effect comes about due to flat-fielding, a process which
presumes that all pixels are the same size on the sky. If, for
example, a given pixel is smaller on the sky than average, then
dividing by a flat-field artificially boosts the flux of that
pixel. In large mosaic cameras, which occupy a significant fraction of
the telescope field and hence have significant distortions, this
variation in pixel size must be corrected by normalizing to the pixel
area. This is done at the catalog stage for the RCS - unlike in
dithered data in which these corrections must be applied to the image
directly prior to stacking the images.

The second step of the astrometric calibration maps the camera
coordinates onto the sky, again using a match to the USNO-A2.0
catalog. This is done by establishing the nominal center position of
the field by visual inspection, followed by an automatic, iterative
algorithm which repeatedly matches objects in position and magnitude
simultaneously until a satisfactory mapping is achieved across the
whole image. Typically 300-500 stars per pointing are used to
establish this final astrometric solution. The final accuracy of the
astrometric solution is indicated by the residuals between the
positions of objects in the RCS and the USNO-A2.0. Typically these
residuals are 1/3rd of an arcsecond per coordinate, consistent with
being dominated by uncertainties in the USNO-A2.0 catalogs. An
excellent secondary check of the accuracy of the astrometric solutions
is the comparison of positions in the RCS photometric calibration
fields to the equivalent SDSS early release data. The comparison
between RCS and SDSS positions on three separate Landolt fields is
shown in Figure 6. After subtraction of a systematic offset of a few
tenths of an arcsecond, the agreement is very good. As shown in Figure
6, this comparison is consistent with an uncertainty between the SDSS
and RCS positions of about 140 milliarcseconds per coordinate. The
total estimated astrometric uncertainty in the SDSS data is on order
of 50-100 milliarcseconds per coordinate \citep{sto02}; Figure 6
suggests that the RCS positions for bright but unsaturated objects are
uncertain at a similar level.

Once the astrometric solution has been established for each camera
pointing, the data must be stitched together to form a master patch
catalog. This is done simply by locating the edge of each pointing and
then taking the midpoint between the edges of it and the adjacent
pointings as the boundary of each pointing. Data outside this boundary
are clipped. The boundary also has a two arcsecond width, to ensure that
objects near the edge of a given pointing do not re-occur across the
boundary in data from the adjacent pointing. This results in a
negligible loss of area and is insignificant in comparison to the
inter-chip gaps in the camera. Currently, the basic data contained in
the master catalogs for each object are: position, $z'$ magnitude,
$R_C-z'$ color, magnitude and color errors, the original chip
coordinates and pointing and chip designations. We expect to add
morphological information once this analysis is fully integrated into
the data pipeline.

During this final stage, we also create a number of ancillary data
products to enable various subsequent analyses. The primary product is
a set of random photometric catalogs, useful at later stages for such
things as generating random catalogs for correlation analyses. These
catalogs, which include photometric information, are made by taking
photometric data from other pointings with similar depths and
assigning random positions to the objects, in the raw chip
coordinates. These data are then assembled analogously to the real
master patch catalog, using the same inter-pointing boundaries. We
also produce a much larger random catalog of positions only, where
each chip is populated by $\sim$3.5$\times$10$^5$ objects at
semi-random positions, with a modification to the density of points
over the chip to account for the varying pixel size in the mosaic
camera. This catalog is useful for estimating the contribution of
individual chips and pointings to global statistics, in cases where
the area rather than the number of objects (and hence image depth) is
of concern. In detail this catalog is made by placing points on a
nominal grid in which each grid cell has an area of one square
arcsecond. A single point is placed at a random position within each
grid cell, producing a catalog which has a white spatial power spectrum
at fine scales (and so reasonably samples chip edges and other area
cutouts) but does not have signifcant power on large scales (and hence
is less noisy than a true random catalog when used to compute area on
these larger scales). For example, this semi-random position catalog
is used in \S6.2. (deriving cluster richness) to compute the
effective area of the survey in arbitrary regions by Monte Carlo
methods, with proper accounting for chip gaps and survey edges.

\section{Data for Patches RCS0926+37 and RCS1327+29}

The first two RCS patches for which complete data are available are
RCS0926+37 and RCS1327+29. Both were imaged during Run 1-a, Run 2, and
Run 4. The data sets used for these patches, as previously noted, are tabulated
in Tables 3 and 4.  The data used here were processed using the
pipeline described above. The final assembled patches cover areas of
5.59 and 4.54 square degrees, respectively.

Figure 7 shows the $R_C$ galaxy counts from the two patches. As
expected, the galaxy counts are very similar, with deviations at the
faint end due to differences in image depths between the two
patches. From this, it is evident that the typical 100\% completeness
depth for galaxies is about $R_C\sim24.0$. Based on extensive past
experience, this corresponds to about a 10$\sigma$ limit for point
sources. The target depth for these data was $R_C$=24.6 at a 5$\sigma$
limit for point sources; the measured value is $R_C\sim24.7$,
consistent with expectations. The variation in depths on a
chip-to-chip basis for both filters is shown in Figure 8, where the
5$\sigma$ point source limit has been estimated empirically for each
chip. Results are shown for both normal and red-sensitive
chips. Extensive tests using similar data \citep{yee91} demonstrate
that the 100\% completeness limit for galaxies is typically 0.7-0.8
magnitudes brighter than the 5$\sigma$ point source limit, completely
consistent with Figures 7 and 8.

Figure 9 shows two color-magnitude diagrams for patch
RCS0926+37 - one for a random set of 20,000 stars and the other for a
random set of 20,000 galaxies. The markedly different distributions
show that the star-galaxy separation works very well.

\section{Cluster Finding with the RCS Data}
The basis of the cluster finding algorithm used to find clusters in
the RCS data is given in \cite{gla00}. In brief, this algorithm works
by searching for density enhancements in the four dimensional space of
position, color and magnitude. In practice, the algorithm works by
cutting up the color-magnitude plane for galaxies into a number of overlapping
color slices, corresponding to expected cluster red sequences over a
range of redshifts. For each slice, the magnitude weighted density of
galaxies is computed using an appropriate smoothing kernel. The
density values are translated into gaussian sigmas by comparison to
the distribution of background values in bootstrap maps. The
individual slices are assembled into a datacube which is then searched
for significant peaks.

\subsection{Algorithm Modifications}

We have applied a modified version of the \cite{gla00} algorithm to
the RCS data for patches RCS0926+37 and RCS1327+29. The smoothing
kernel used is as described in \cite{gla00}, with a core radius of
300 $h^{-1}$ kpc, and the algorithm is in most details identical. The
two significant changes are an enhancement to the algorithm designed
to account for pointing-to-pointing and chip-to-chip variations in RCS
images, and a new algorithm for selecting peaks in the final
datacube. Each modification is described in detail below.

\subsubsection{Algorithm Enhancments to Account for Variations in Survey Data}

The complications due to image variations are subtle, and must be
carefully accounted for. The basic problem is that the sampling depth
on any one image from a single chip varies in both filters from all
other single chip images in the survey. The depth is a function of the
QE of the chip in each filter (this is rather varied in the $z'$
filter, particularly because of the two chip types in the camera)
combined with the sky brightness and seeing for each
pointing. Moreover, the depths achieved are a function of source type
- for example, poor seeing will affect the depths achieved for point
sources more than for large galaxies, whereas a brighter sky will have
less of an impact on point sources compared to larger galaxies. Also,
the QE variations are chip-to-chip (in that the same relative
sensitivity is preserved between different chips in the mosaic
regardless of sky brightness and seeing variations), and the seeing
and sky brightness variations affect the depths on a
pointing-to-pointing basis.

A trivial way to correct for these depth variations \citep[which was
used in][in application to the CNOC2 data]{gla00} is to simply cut the
entire photometric catalog to the depth of the shallowest image in the
survey. Clearly, for a large-scale survey this is a non-optimal
approach. Also, unless the cuts are made at rather bright limits this
approach does not even work well, because the variations in the photometric
uncertainty at the faint end of the distribution still produce
statistical differences between different regions of the survey, even
when all portions of the survey have the same nominal completeness.

We have extensively explored the idea of using the photometric
catalogs with random positions (described in \S4.3.4) to normalize
the density maps.  The matching random-position photometric catalogs
for each chip are generated by drawing photometry from chips of
similar depth coupled with random positions and provide data which
are in principle statistically similar to the actual data.  Nominally,
one could then use these data to re-normalize in some way the actual
data to account for depth variations. In practice we find that this
approach is insufficient, in that residual chip-to-chip variations are
still evident in the density maps produced by the cluster-finding
algorithm.

In order to fully correct these chip-to-chip and pointing-to-pointing
variations, we instead use a strategy based on sampling of the actual
data. Consider first that each full patch typically consists of a
total of 180 (i.e., 15 pointings times 12 chips) individual pairs of
$R_C$ and $z'$ images, each of which has a slightly different sampling
depth. If each single chip image were totally isolated on the sky, one
could in principle trivially use bootstrap re-samplings of only that
one image (technically, bootstrap re-samplings of the corresponding
catalog) to compute the significance of peaks found in the various
density maps arising from the different color slices of that
catalog. However, apart from the fact that the individual images are
not isolated on the sky, this approach is not feasible, since a single
large cluster can dominate the signal on a given chip, at least in the
color slices corresponding to the cluster's redshift. In practice
then, one must isolate regions larger than a single chip which have
similar statistics from which one can estimate the significance of any
given peak. Thus, for each pixel in a density map, the goal is to
locate a subset of all pixels in the map which sample regions with
similar statistical properties; the same pixels from a large set of
bootstrap re-sampled maps then provide the necessary background
distribution used to compute the significance of the measured value.

Following \cite{gla00}, recall that the kernel-smoothed density map of
a given color slice is an array of $n\times m$ pixels, encoding the
kernel-smoothed density value, $\delta_{ij}$, at the corresponding
location. In practice, because the camera is a tightly packed mosaic,
and because the pointings overlap, the smoothing kernel centered at a
given location often spans multiple chips or pointings, and so the
measured $\delta_{ij}$ at that point is typically influenced by
several datasets each with slightly different depth. The value of
$\delta_{ij}$ at any given location is then a reflection of the local
density of galaxies of a given color at that position, and the
sampling depth of the datasets which contribute to that measured
density. To establish which other pixels in the maps to use as
backgrounds, we want to somehow average over the region of the dataset
corresponding to each particular observation, and then use only pixels
of similar sampling depths (indicated by having a similar average
value of $\delta_{ij}$) when computing the significance of the
measured $\delta_{ij}$s. There are likely two partitions of the data
in a given patch which are significant - one corresponding to the
individual pointings (nominally 15 regions per patch) and one
corresponding to the individual chips (nominally 12 regions per
patch). In principle we are thus interested in deducing a total of 180
``average values'' from the input map of $\delta_{ij}$s, where each
average value corresponds to a given chip and pointing combination, and
can be estimated from a region of the input map. In practice this
number is often larger since data from each run is considered separately
due to significant changes in the instrument from run to run.

To find the appropriate portion of the input map from which to
estimate a given average value, we turn to the random position-only
catalog described in \S4.3.4. In this catalog each chip on each
pointing is represented by a semi-random distribution (in the original chip
pixel coordinates) of $\sim3.5\times10^5$ points with no associated
photometry. These positions are run through the same master catalog
assembly process, and produce a position-only master catalog which
precisely reproduces the overlap cuts which are used when stitching
the pointings together. The contribution of any one chip to the map
can be estimated from this catalog simply by applying the same density
estimator used in creating the real-data density map to the random
points which come from only that chip. We use this random catalog, in
conjunction with the actual input catalog, to estimate the ``average
value'' of each pixel in the input map (call the input density map
$M1$). This is done in a two discreet stages.

First, we want to produce a map, called $M2$, which represents the
average value of each pointing in $M1$. To do this, we consider each
pixel in $M1$ as a measure of the sum of a contribution from each
pointing. Algorithmically, we construct a highly over-determined set of $i \times j$ linear
equations (one for each pixel) in which we presume that each
measured $\delta_{ij}$ is given by
\begin{equation}
\delta_{ij}=\sum_{P=1}^{15}W^P_{ij}A^P,
\end{equation}
\noindent
where $W^P_{ij}$ is the weight (calculated from the position-only
catalog) of the $Pth$ pointing at pixel $i,j$, and $A^P$ is the
average value of the $Pth$ pointing. The $A^P$s (fifteen of them for a
typical full patch) are unknowns.  We solve this using singular-value
decomposition and hence recover optimal estimates for the $A^P$s. These
$A^P$s are then used to create a density map, via Equation 2, of
$\delta_{ij}'$s, which is a map of the pointing-averaged value at each
pixel. This is the desired map $M2$. $M1$ is then divided by $M2$ to
produce a map (call it $M1'$) which is devoid, on average, of
variations on a pointing-to-pointing basis.

Figure 10 illustrates the generation of the map $M2$ and $M1'$ in more
detail. The left panel shows the input map $M1$, in this case
corresponding to a color slice for the $z$ range 0.555-0.575 in patch
RCS1327+29. The center-left panel shows an example map of W$^P_{ij}$, in
this case for the A2 pointing. Nominally 15 such maps are generated
from the position-only random catalogs, in order to set up Equation
2. The center-right panel shows the resulting map $M2$, and the right
panel shows the map $M1'$.  

We next apply a similar procedure to $M1'$ in order to produce the
average value of each chip in $M1'$ (we will call this map $M3$). In
this case, ``chip'' refers to all occurrences of a given chip of the
mosaic within the entire patch. In practice, this means solving for a
number of further unknowns akin to the $A^P$s in equation 2. These
unknowns correspond to chips from each run contributing to the patch;
each run is treated independently because the camera used underwent
continual refits during the course of the survey. We again solve for
these average values using singular-value decomposition of a highly
over-determined set of $m\times n$ linear equations, and from this
deduce the map $M3$. 

The product of maps $M2$ and $M3$ yields a map, call it $M4$, which
gives the average value of the $\delta_{ij}$s across each chip on
each pointing in $M1$, where each average has been estimated over an
area significantly larger than a single chip (solving the problem of a
large cluster dominating in a small region) and each averaged pixel
properly accounts for the relative contributions of all chips and
pointings.  From this, each value of $\delta_{ij}$ in $M1$ can be
transformed into a probability by using only those portions of the
bootstrap background maps which have similar average values in
$M4$. This ensures that the significance of any given peak is assessed
only in comparison to data of similar depths.

Similarly to Figure 10, Figure 11 illustrates the generation of the
maps $M3$ and $M4$. The input map, $M'1$, is the rightmost panel of
Figure 10.  The left panel of Figure 11 shows an example map of
W$^P_{ij}$, in this case for chip 2 in the data corresponding to Run 2
(for a total of ten occurrences of this chip in this particular patch -
see Table 4). There are total of 22 such maps for this patch, since
there are data from both Run 1-a (ten chips) and Run 2 (twelve
chips). The resulting map of average values across chips, $M3$, is
shown in the center-left panel. The map $M4$ (the product of maps $M2$
and $M3$) is shown in the center-right panel. The right panel shows
the map $M1$(see the left panel of Figure 10) divided by $M4$. Note that
any apparent structure on both chip scales and pointing scales is now
negligible.

\subsubsection{Identifying Cluster Candidates}

The other major change from the algorithm developed in \cite{gla00} is
the method used to find peaks in the datacube of $\sigma_{ij}$s. In
\cite{gla00} we used the readily available three dimensional peak
finding algorithm of \cite{wil94} to select significant peaks in the
datacube, and to separate nearby peaks. Further experimentation has
shown, however, that the separation of nearby peaks is better
accomplished using a more physically motivated model, and so a special
purpose method was developed.

The final peak-finding algorithm is relatively simple. Peaks are
identified by finding the highest-valued voxel in the datacube, and
then iteratively connecting all adjacent voxels down to a chosen
threshold. This process is iterated, ignoring all previously flagged
voxels, until all ``significant'' peaks are flagged. The significance
cut is an arbitrarily chosen value which attempts to balance
completenes and contamination in the catalog. In this paper we use a
cut of 3.29$\sigma$ (corresponding to a nominal 1 in 1000 chance of
random occurrence). Each peak above this level is traced down to a
lower threshold of 2.5$\sigma$.

This simple-minded approach can and does connect subpeaks which appear
somewhat separated in the datacube. To investigate the physical
significance of these peaks, we have developed a modified algorithm
capable of breaking up a single region connected at a relatively low
threshold into its constituent sub-peaks. Figure 12 plots the angular
and redshift separations of all possible pairs of peaks identified by
this modified analysis, applied to the patch RCS0926+37. Two sets of
values are plotted: those which correspond to pairs of subpeaks which
are within a single primary peak, and those which correspond to pairs
of separate primary peaks.  Both in angular coordinates and redshift,
this provides a natural separation in scale. Notably, for the angular
separations the dividing region between the two scales is close to the
expected virial radius for clusters. Furthermore, in redshift space
the division appears to correspond to the expected redshift
uncertainty for individual clusters at all but the highest
redshifts. At the highest redshifts individual peaks are at a
generally lower signal-to-noise ratio, and redshifts may be more
systematically uncertain than simple models indicate (see \S6.2.2),
and the excess difference between sub-peaks at these extreme redshift
is likely not significant. Thus, in almost all cases, connected
subpeaks are in fact closer in projected separation than the size of a
single cluster, and generally indistinguishable in redshift. Moreover,
separate primary peaks are almost never so close. We thus choose, on
reasonable physical grounds, to call all such connected subpeaks a
single cluster.

\subsubsection{Uncertainties in the Significance of Cluster Candidates}
The significance of a given cluster has some associated uncertainty,
which derives in part from computational limits. This computational
uncertainty is separate from uncertainty in the significance (and
other derived quantities such as redshift) which derives from, for
example, photometric uncertainty in the input data. The latter is
difficult to quantify in the absence of repeated imaging of the same
area of sky, but is likely to be small for most cluster candidates
since the cluster signal is an aggregate from many objects. The
computational uncertainty can be readily quantified however, and is
primarily due to the limited number of bootstrap realisations which
can be reasonably computed. Consider, for example, that a
$\sim$5-sigma peak is roughly a one in one million event. Thus, to
make a measurement of a 5-sigma peak at a signal-to-noise ratio (S\/N)
of ten requires about 100 million bootstrap samples. If one uses only
a tenth of each bootstrap map to compute the significance of any given
pixel in the input map (i.e., as described in detail in \S6.1.1) then
about 10$^9$ bootstrap map pixels are needed to describe each possible
5-sigma peak at a S\/N of ten. These computational limits motivated
\cite{gla00} to use a fitted version of the distribution of
$\delta_{ij}$'s in the bootstrap maps, in attempt to suppress noise in
the final significance maps. We use a similar procedure here, but apply
a fitting function to only the high-valued end of the distribution,
where the bootstrap maps are insufficiently well sampled. Below
approximately 4-sigma, the actual distribution of bootstrap values is
used to compute significance, and above about 4.5-sigma the fitted
distribution is used, with a transition region in between these values
where a weighted mean of both is used.

Figure 13 shows an estimate of the uncertainty in the significance of
voxel values in the datacube for patch RCS0926+37. This has been
computed by comparing the voxel values across different runs of the
cluster-finding algorithm. Note the general increase in uncertainty
toward higher values of sigma; the use of a fitting funtion to the
bootstrap distributions suppresses pixel-to-pixel noise in a given
slice in a particular datacube, but does not suppress the noise across
different runs of the bootstrap analysis. It is important to note in
this analysis that the computational uncertainty in sigma at any given
value of sigma is always much less than the difference between the
values of sigma and the $\sim$ 3-sigma threshold used to define the
catalogs, and that at this threshold the uncertainty in sigma is
extremely small. Thus in the catalogs described below, the uncertainty
in sigma has no significant effect on the inclusion of objects in the
catalog.

\subsection{Two Catalogs of RCS Clusters}
\subsubsection{The Catalogs}
Tables 5 and 6 give the cluster catalogs for patches RCS0926+37 and
RCS1327+29 respectively, ordered by redshift, to a significance cut of
3.29$\sigma$. The peak significance of each cluster is given. Each
cluster is identified with a name of the format RCS~J{\it
HHMMSS+DDMM.m}, with coordinate values truncated, as suggested by IAU
nomenclature conventions. Clusters are listed only at redshifts greater
than 0.20. The redshift accuracy at lower redshifts is compromised by
the RCS filters. We are in the process of integrating complementary
$B$- and $V$-band data in to the RCS databases; once available these
data will be used to define a lower redshift complement to the
catalogs presented here. Precise positions for each candidate are
provided in J2000.0 coordinates. The final positions are found using
an iterative centroiding algorithm using a three dimensional gaussian
kernel applied to the datacube. The centering kernel has a spatial full
width at half maximum of 250 $h^{-1}$ Mpc, and has a sigma width in
redshift of 1.5 voxels. The voxel with the maximum value within the
identified peak is used as the starting point for centering.

We also provide in Tables 5 and 6 the offset in arcseconds between
this final position and the position of the brightest cluster galaxy.
This galaxy is selected by considering all galaxies which are interior
to the projected $2.5\sigma$ boundary in the datacube which defines
the cluster candidate, and which have colors within
$\sqrt{0.2^2+\Delta C^2}$ magnitudes of the expected red sequence
color at the cluster redshift, where $\Delta C$ is the color error of
each galaxy. The minimal cutoff of 0.2 magnitudes in color is the same
as that typically used to separate blue and red cluster galaxies
\citep[e.g.,][]{but84}. Each galaxy considered is assigned a score
equal to its $z'$ magnitude, minus the value of $\sigma_{ij}$ at that
line of sight at the cluster redshift. The lowest ranking object is
picked as the nominal brightest cluster galaxy. The use of weighting
by the $\sigma_{ij}$'s, in addition to simply considering the
magnitudes, guards against unassociated bright objects on the
periphery of the cluster being selected as the center. Large values of
this offset between the position of this galaxy and the cluster center
may indicate an incorrect central galaxy, or a cluster with a
filamentary shape or ill-defined center (such as a double cluster).

In addition to positions for each cluster, Tables 5 and 6 give
estimates of the apparent projected size and shape of each
cluster. These values are derived by considering all voxels
corresponding to the cluster in the $\sigma_{ij}$ datacube, projected
along the redshift axis. A size and ellipticity is computed by
considering weighted moments of this projected distribution, with each
input voxel assigned a weight equal to its value in excess of
$2.5\sigma$. Tables 5 and 6 give the resulting ellipticity, and the
size of the semi-major axis in arcseconds. Clusters with unusually
large sizes or ellipticities are likely multiple associated
structures. As a final diagnostic, Tables 5 and 6 also provide an
estimate of the redshift ``range'' for each cluster, where the range
is found from the minimum and maximum redshifts ascribed to the set of
voxels which make up the cluster peak in the $\sigma_{ij}$
datacube. As in the projected size and shape, egregiously large values
of this range in comparison to other clusters of similar significance
and redshift may indicate a projection of some sort. In any obvious
cases of projections or double clusters we resist the temptation to
modify objects individually, preferring instead to define a catalog
based strictly on a single algorithm. This facilitates automated
comparison to future modeling efforts.

The expected false-positive contamination rate for the combined
catalog is discussed extensively elsewhere \citep{gla02a}.  Projection
effects due to the clustering and random projection of nominally field
galaxies is expected to be less than 5\% at all redshifts, consistent
with that seen in an empirical test using a combination of photometric
and redshift data from the much smaller CNOC2 Survey \citep{gla00}. A
larger fraction of all clusters in Tables 5 and 6 will have some amount
of projected struture; these are real clusters, but their apparent
properties may be modified by projection of galaxies from nearby
clusters and groups in the cosmic web. Such cases of associated
multiple structures are in part distinguishable by the size and shape
criteria outlined above. 

Figure 14 provides examples color-magnitude plots for representative
clusters from Tables 5 and 6. In some cases, particularly for lower
significance systems, the red sequence is not overwhelmingly apparent
to casual examination. For such systems it is the aggregate signal in
color, magnitude and position which results in a detection, and direct
visual examination does not always yield similar confidence.

Figures 15a-15bi show color images of each clusters listed in Table
5, for the patch RCS0926+37. Figures 16a-16au similarly show all
clusters in Table 6 for the patch RCS1327+29. Each figure provides
color images of four clusters, constructed from the $R_C$ and $z'$
survey images. These are overlaid by a contour map of the projected
$\sigma_{ij}$ map, as used above to estimate the cluster size. For
each cluster a larger scale version of this map is also shown, which
give a visual indication of possible nearby clusters, many of which
will themselves be listed in Tables 5 and 6.

\subsubsection{Redshift Calibration and Uncertainties}
The red-sequence model used for the cluster finding presented here is
the $z_{f}=2.5$ model described in \S2.1. The model has been
fine-tuned by adjusting the color to match the redshifts of several
known clusters in the patch RCS1327+29. The required color adjustment
is a few hundredths of a magnitude, well within the expected
uncertainties between absolute calibration of the photometry and the
modeling. Initial spectroscopy of a subset of RCS clusters at
redshifts $0.2<z<1.0$ shows that the redshifts derived from the
photometry are typically accurate to better than 0.05 over this
redshift range \citep{gla03}, and possibly as good as 0.03 in fields
with optimal photometric calibration. Extensive modeling of the RCS
data \citep{gla02a} confirms this expected \citep{gla00} result, and
suggests that redshift errors should increase at the lowest redshifts
and at $z>1$. At the highest redshifts, the 1-sigma uncertainty in the
photometric redshifts is approximately 0.1, due to a combination of
poorer sampling of the 4000\AA~break by the $R_C$ and $z'$ filters,
and larger photometric uncertainties on the increasingly faint and red
cluster galaxies. At the highest redshifts there is also a fundamental
ambiguity regarding the appropriateness of the particular red-sequence
model used, in that changes in the details of the model produce
significant changes in the expected colors of the cluster red sequence
at $z>1$ \citep{gla03}. Additionally, the near-degeneracy between
$R_C-z'$ color and redshift at $z\sim1$ (see Figure 1) tends to scatter
clusters at $z\sim1$ to either slightly higher or lower redshifts, more
so than at other redshifts, which results in an apparent depopulation
of the cluster population at that redshift.

\subsubsection{Richness Estimates}
Tables 5 and 6 also provide a richness estimate for each cluster. The
clusters in the catalogs are characterized by the richness parameter
\bgc, the amplitude of the cluster-center-galaxy correlation function
computed individually for each cluster assuming a distribution of
excess galaxies of the form $\xi(r)\sim B_{gc} r^{-1.8}$ (see Yee \&
L\'opez-Cruz 1999 for a detailed discussion of the derivation and
properties of the parameter).  The \bgc~parameter has been found to be
a robust richness estimator \citep[][and references therein]{yee99},
and correlates with important cluster attributes such as velocity
dispersion, mass, and X-ray temperature and luminosities for a set of
X-ray luminous clusters with scatters of 15 to 40\% \citep{yee02}.
                                                                                                            
For the RCS clusters, we compute \bgc~using a more refined method than
that in Yee \& L\'opez-Cruz by fully utilizing the two-band
photometric data.  Galaxies are counted in regions defined in the
color-magnitude diagram (CMD) to minimize projection effects and
counting uncertainty.  This is essential for high-redshift clusters,
as the effect of projected galaxies is substantially more serious than
that for lower-redshift clusters.  A fiducial color-magnitude relation
for each cluster is defined by that used to find the cluster in the
first place, from which regions in the CMD for galaxy counting are
established.
                                                                                                            
We compute two different \bgc~parameters for each cluster: one using
all excess galaxies (deriving what we call the total \bgc, or
{$B_{gcT}$), and the other using the excess red-sequence galaxies
(deriving the ``red-sequence'' \bgc, or $B_{gcR}$).  In the former we
count galaxies in the CMD bounded in colors by: 0.20 mag in $R_C-z'$
to the red of the red-sequence relation, and 0.25 mag in $R_C-z'$
to the blue of a blue star burst SED (colors for this limit are taken
from a GISSEL \cite{bru93} pure starburst model of intermediate
metallicity); and bounded in magnitude by 3 mag brighter than the
expected $M^*$, and 2 mag fainter (or the 100\% completeness limit
magnitude, whichever is brighter).  Typically, the depth of the
photometry allows the sampling (in the $z'$ band) to 2 mag below $M^*$
to a redshift of $\sim0.65$.  At higher redshift, the \bgc~parameter
(which is normalized by the galaxy luminosity function) is derived
using counts to a shallower effective depth, which increases the
uncertainty, especially at $z>1$ where on average we sample to less
than 1 mag below $M^*$ (see the detailed discussion in Yee
\&~L\'opez-Cruz 1999).  The galaxy counts are done over an aperture of
radius 0.5 $h_{50}^{-1}$ Mpc, centered on the nominal cluster center
(see \S 6.2). We retain h=0.5 to keep consistency with \bgc~~
measurements from \cite{yee99} and \cite{yee02}. Galaxy counts are
modified by the sampling area as appropriate, with the sampling area
estimated by examining the semi-random position-only catalogs (see
\S4.3.4).

 For clusters with red-sequence photometric redshifts of less 0.45, we
use the $R_C$ catalogs for galaxy counting, while for clusters with
$z>0.45$, the $z'$ data are used.  At $z\sim0.45$ the blue part of the
$R_C$ band begins to encroach on the 4000\AA~break; hence using the
$z'$ data diminishes uncertainties in both galaxy evolution and
K-correction that are needed to apply to the photometric data for
proper galaxy counts.  We use the $R_C$-band luminosity function in
\cite{yee99} for the normalization of the galaxy counts, although we
note that the simple parameterisation of the evolution in $M^*$ is
almost certain not to be correct for blue galaxy dominated clusters at
high redshifts.  The K-corrections used are based on galaxy spectrum
models from \cite{col80}.  We correct the LF to $z'$-band using
$R_c-z'$ colors from the fiducial red-sequence model discussed in
\S2.1.  We note that we obtain very similar \bgc~values, well within
the uncertainties, for clusters at $z\sim0.45$ when they are computed
using both $R_C$ and $z'$ band data.

The average background galaxy counts used to perform the statistical
count corrections are obtained directly from the very large amount of
survey images themselves, ensuring total self-consistency.  For each
cluster, the average count is derived using $\sim$10 square degrees of
the RCS data (both patches in their entirety) with identical color and
magnitude cuts as those used for the cluster.  Furthermore, the large
area available also allows us for the first time to derive the
estimate in the stochastic variation in the background counts entirely
empirically.  The variance in the background count is derived for each
cluster by  randomly sampling the counts in several hundred areas with
the same angular size and color and magnitude cuts as those used to
compute the \bgc~value.  This is incorporated into the uncertainty
estimate of the richness parameter.

The highly uncertain evolution of cluster galaxies at $z>0.5$
motivates the use of a red-sequence \bgc.  The red galaxies in
clusters are much slower evolving and likely to be largely in place
even at redshift one.  Thus, using only the red galaxies as an estimate
to the richness will provide a much more stable measurement, less
affected, for example, by the varying blue fraction of clusters.  (We
note that in our preliminary measurements the blue fractions vary from
$\sim$ 0.2 to 0.8 for clusters at the high end of our redshift range.)
This in turn should allow us to obtain more stable estimates of
cluster properties such as velocity dispersion, mass, X-ray
temperature and luminosity based on (modified) calibrations such as
those from Yee \& Ellingson (2003).  A more detailed discussion of
$B_{gcR}$ will be given in future analysis papers on the RCS sample.
We compute $B_{gcR}$ by using a color slice with an upper (red) color
bound that is 0.2 mag redder than the fiducial red-sequence, and a
lower (blue) color bound equivalent to the mid-point in color between
an elliptical and Sbc galaxy at the relevant redshift.  The
background counts are obtained using the identical color cuts. The two
\bgc~parameters are listed for each cluster in our catalogs in Tables
5 and 6.

Figure 17 shows the relationship between the red-sequence richness,
$B_{gcR}$, and the detection significance. There is a broad
correlation between the two, with significant scatter. There are
numerous reasons to expect significant scatter in this correlation;
for example the richness is insensitive to the cluster concentration
due to the large aperture over which it is measured, whereas detection
significance can be significantly boosted if a cluster is particularly
compact. The depth of the data also affects the detection
significance rather strongly, whereas the richness is, by design,
constructed to guard against such effects.

%We
%also use the position-only random catalog over the same region as each
%cluster to make small corrections to the computed excess and
%background galaxy counts. This process (essentially a Monte-Carlo
%estimation of the actual area enclosed by the counting aperture)
%corrects for incompleteness due to interchip gaps and patch edges.

Figure 18 summarizes the richness and redshift distribution for the
two patches in aggregate. Redshift distributions are shown for four
$B_{gcR}$ richness ranges corresponding to traditional Abell Richness
Classes (ARC), as calibrated in \cite{yee99}, without modification to
account for the color cuts used here.  As expected, the bulk of the
clusters are of ARC 0 and poorer, and some fraction of these poorest
systems are likely best termed groups. At higher redshifts the sample
has proportionately more richer systems, again as expected since the
poorer systems fall out of the sample due to incompleteness.
 
\subsubsection{Known Clusters}
The patch RCS1327+29 overlaps survey fields from both the Palomar
Distant Cluster Survey \citep{pos96} and the older survey of
\cite{gun86}. There are 4 spectroscopically confirmed
clusters\footnote{As recorded in the NED database.} from these prior
surveys within the boundaries of the RCS patch which we consider of
sufficient reliability to warrant discussion here. All are present in
Table 6, and 3 of these 4 clusters are detected at greater than 4
sigma. Table 7 lists each cluster, along with the RCS redshift
estimate and detection significance. In all cases the estimated
redshifts are consistent with the spectroscopic redshifts available
from the literature.

\section{Summary}

The RCS, now complete, is the largest moderately deep two-filter
imaging survey done to date. We have developed an extensive processing
pipeline to handle the data flow from this survey, which has been
described extensively. Through judicious choices of observing
strategies, the data are readily handled using essentially standard
methods, modified slightly to accommodate the peculiarities of mosaic
cameras. We have paid particular attention to the uniformity of the
photometric calibration of the survey, since this affects the accuracy
of cluster redshifts. The typical systematic uncertainty in the colors
is expected to be less than 0.03 magnitudes in most cases,
corresponding to systematic redshift uncertainties which are no more
than the expected random errors at all redshifts.

The complete data processing pipeline has been applied to data from
the first two completed RCS patches. Analysis of the basic data
products such as galaxy counts illustrate that the catalogs contain no
major systematic uncertainties. The targeted depths have been reached
in both filters; the measured typical 5-sigma point source limits are
23.8 and 24.9 in $z'$ and $R_C$ respectively.  

Included in this paper are two catalogs of clusters at $z>0.2$, from
the patches RCS0926+37 and RCS1327+29. Alone these two catalogs
represent a significant increase in the total population of known
clusters and groups, particularly at $z>0.5$. Processing of the RCS
data is being finalized, and we expect to release further cluster
catalogs soon \citep[e.g.,][]{bar02}, with an eventual full release
of all survey data (processed images, and both primary and derived
catalogs).

\acknowledgments

We are grateful to the CFHT Canadian TAC for the generous allocations
of CFHT time which made this project possible. We are also thankful
for the efforts of the CFHT staff in providing the CFH12K camera; in
particular we thank J.C. Culliandre, whose assistance has been
invaluable. M.D.G. acknowledges financial support from the Canadian
Natural Sciences and Engineering Research Council (NSERC) via PGSA,
PGSB and PDF Fellowships. This project is supported in part by an
NSERC operating grant and a University of Toronto grant to H.K.C.Y. We
also thank a number of summer students, Helen Kirk, Danica Lam, and
Trevor Evans, for assistance with running the photometric pipeline and
visual verification of the object-finding. Finally we would like to
thank the anonymous referee, whose detailed comments served to greatly
improve the clarity of this paper.

\clearpage
%\onecolumn

\begin{table}
\centering
\begin{tabular}[!hb]{crrcccr}
\hline
Patch& RA(2000)& DEC(2000)& 100$\mu$ & $E(B-V)$& Area (deg$^2$) & Notes \\
\hline
0226+00& 02~26~07.0& +00~40~35&2.73&0.036&4.81&CNOC2 Patch\\
0351--09& 03~51~20.7& --09~57~41&2.57&0.043&4.79&\\
0926+37& 09~26~09.6& +37~10~12&0.39&0.012&5.59&CNOC2 Patch\\
1122+25& 11~22~22.5& +25~05~55&N/A&0.018&4.78&\\
1327+29& 13~27~41.9& +29~43~55&-0.02&0.012&4.54&PDCS Patch\\
1416+53& 14~16~35.0& +53~02~26&-0.20&0.010&4.53&Groth Strip\\
1449+09& 14~49~26.7& +09~00~27&0.89&0.029&4.17&CNOC2 Patch\\
1616+30& 16~16~35.5& +30~21~02&1.03&0.038&4.26&\\
2153--05& 21~53~10.8& --05~41~11&3.04&0.035&3.43&CNOC2 Patch\\
2318--00& 23~18~10.7& --00~04~55&2.96&0.057&4.84&\\
\hline
\end{tabular}
\caption{Basic Data for the RCS CFHT Patches} {\footnotesize IRAS
100$\mu$ brightnesses are given in mJy~ster$^{-1}$, and $E(B-V)$
estimates are given in magnitudes. In all cases these values are the
median of all measurements in the patch. Patch names and positions are
from the coordinates of the pointing B3, which is the central pointing
for a standard patch.}
\end{table}

\begin{table}
\centering
\begin{tabular}[!hb]{lrl}
\hline
Run&Dates&General Comments\\
\hline
Run 1-a& May  5-6, 1999 & photometric, excellent seeing\\
Run 1-b& July 7-9, 1999 & shared, 1 night of 3, photometric\\
Run 2  & January 7-14, 2000& shared, 4 nights of 8, mostly photometric\\
Run 3  & July 23-26, 2000& 1 night lost to weather, partially photometric\\
Run 4  & January 27-28, 2001& mostly photometric\\
\hline
\end{tabular}
\caption{Basic Data for the RCS CFHT Runs}
\end{table}

\begin{table}
\centering
\begin{tabular}[!hb]{cccc}
\hline
Pointing&Run&Seeing($''$)&Int. Time(sec)\\
&&$R_C$/$z'$&$R_C$/$z'$\\
\hline
0926A0&4 &1.03/1.05&900/1200\\
0926A1&2 &0.87/0.74&840/1200\\
0926A2&2 &0.81/0.99&900/1800\\
0926A3&2 &1.00/1.02&840/1140\\
0926A4&2 &0.97/1.16&840/1140\\
0926A5&2 &1.11/0.99&840/1140\\
0926B0&4 &1.09/1.05&900/1200\\
0926B1&2 &0.76/0.68&900/1200\\
0926B2&2 &0.77/0.73&900/1200\\
0926B3&2 &0.82/0.68&900/1200\\
0926B4&2 &0.65/0.66&900/1140\\
0926B5&2 &0.89/0.80&900/1200\\
0926C0&4 &1.02/0.85&840/1100\\
0926C1&2 &0.94/0.86&840/1200\\
0926C2&1-a&0.61/0.57&900/1200\\
0926C3&1-a&0.71/0.63&900/1200\\
0926C4&1-a&0.73/0.69&900/1200\\
0926C5&2 &0.97/0.89&840/1200\\
\hline
\end{tabular}
\caption{Observational Data for the Patch RCS0926+37}
\end{table}

\begin{table}
\centering
\begin{tabular}[!hb]{cccc}
\hline
Pointing&Run&Seeing($''$)&Int. Time~(sec)\\
&&$R_C$/$z'$&$R_C$/$z'$\\
\hline
1327A1&2 &0.91/1.04&500/1400\\
1327A2&2 &0.72/0.88&900/1200\\
1327A3&2 &0.77/0.66&840/1140\\
1327A4&2 &0.75/0.70&840/1080\\
1327A5&2 &0.79/0.87&900/1200\\
1327B1&2 &0.78/0.69&900/1200\\
1327B2&2 &0.76/0.67&900/1200\\
1327B3&2 &0.87/0.68&840/1200\\
1327B4&2 &0.91/0.78&840/1140\\
1327B5&2 &1.18/1.10&500/1100\\
1327C1&1-a&0.62/0.55&840/1200\\
1327C2&1-a&0.57/0.59&900/1200\\
1327C3&1-a&0.61/0.54&900/1200\\
1327C4&1-a&0.69/0.60&900/1200\\
1327C5&1-a&0.65/0.59&840/1200\\
\hline
\end{tabular}
\caption{Observational Data for the Patch RCS1327+29}
\end{table}

\begin{deluxetable}{cccrrrrrrrr} 
\tabletypesize{\scriptsize}
\rotate
\tablecolumns{13} 
\tablecaption{Clusters in the Patch RCS0926+37} 
\tablehead{ 
\colhead{Cluster}    & \colhead{RA}    & \colhead{DEC}    &  
\colhead{$z$}    & \colhead{$\Delta z$}    & \colhead{$\Delta$BCG} & \colhead{Size}    &  
\colhead{$\epsilon$} & \colhead{$\sigma_{peak}$}    & \colhead{B$_{gcT}$}    &  
\colhead{B$_{gcR}$}    \\ 
\colhead{}    & \colhead{(J2000)}    & \colhead{(J2000)}    &  
\colhead{}    & \colhead{}    & \colhead{(\arcsec)} & \colhead{(\arcsec)}    &  
\colhead{} & \colhead{}    & \colhead{(Mpc$^{1.8}$)}    &  
\colhead{(Mpc$^{1.8}$)}
}
\startdata 
RCS~J092445+3628.1&09:24:45.70&+36:28:11.6&  0.202&  0.037&  20.6&  44.0&  0.368&  3.58&  495 $\pm$ \phn 187&  393 $\pm$  164\\
RCS~J092414+3710.2&09:24:14.00&+37:10:15.3&  0.220&  0.037& 348.9& 194.0&  0.643&  4.92&  758 $\pm$ \phn 217&  706 $\pm$  200\\
RCS~J092540+3626.8&09:25:40.70&+36:26:49.8&  0.233&  0.037&  25.6& 138.2&  0.225&  6.00&  798 $\pm$ \phn 223&  797 $\pm$  209\\
RCS~J092343+3804.1&09:23:43.00&+38:04:09.6&  0.251&  0.037&   9.9&  33.1&  0.027&  3.77&  199 $\pm$ \phn 164&  178 $\pm$  139\\
RCS~J093004+3828.5&09:30:04.80&+38:28:30.2&  0.262&  0.037& 141.2& 124.6&  0.261&  4.40&  910 $\pm$ \phn 237&  899 $\pm$  220\\
RCS~J093115+3638.4&09:31:15.70&+36:38:24.7&  0.270&  0.037&  11.8&  27.9&  0.026&  3.71&  421 $\pm$ \phn 193&  340 $\pm$  161\\
RCS~J092536+3800.3&09:25:36.50&+38:00:21.6&  0.275&  0.037&  14.6&  56.1&  0.449&  3.55&  501 $\pm$ \phn 201&  369 $\pm$  164\\
RCS~J092523+3708.7&09:25:23.70&+37:08:43.1&  0.275&  0.037&  18.4&  63.7&  0.436&  3.68&  538 $\pm$ \phn 205&  500 $\pm$  179\\
\enddata 
\end{deluxetable} 

\onecolumn
\begin{deluxetable}{cccrrrrrrrr} 
\tabletypesize{\scriptsize}
\rotate
\tablecolumns{13} 
\tablecaption{Clusters in the Patch RCS1327+29}
\tablehead{ 
\colhead{Cluster}    & \colhead{RA}    & \colhead{DEC}    &  
\colhead{$z$}    & \colhead{$\Delta z$}    & \colhead{$\Delta$BCG} & \colhead{Size}    &  
\colhead{$\epsilon$} & \colhead{$\sigma_{peak}$}    & \colhead{B$_{gcT}$}    &  
\colhead{B$_{gcR}$}     \\ 
\colhead{}    & \colhead{(J2000)}    & \colhead{(J2000)}    &  
\colhead{}    & \colhead{}    & \colhead{(\arcsec)} & \colhead{(\arcsec)}    &  
\colhead{} & \colhead{}    & \colhead{(Mpc$^{1.8}$)}    &  
\colhead{(Mpc$^{1.8}$)}
}
\startdata 
RCS~J132757+2900.5&13:27:57.10&+29:00:32.5&  0.201&  0.037&  82.9&  47.8&  0.113&  3.32&  375 $\pm$ \phn 174&  313 $\pm$  154\\
RCS~J132524+2935.9&13:25:24.60&+29:35:57.9&  0.212&  0.037& 111.1& 106.7&  0.550&  3.32&  401 $\pm$ \phn 179&  348 $\pm$  160\\
RCS~J132606+2926.5&13:26:06.70&+29:26:31.2&  0.229&  0.037&  26.8&  52.6&  0.137&  3.96&  422 $\pm$ \phn 185&  303 $\pm$  155\\
RCS~J132523+2839.0&13:25:23.80&+28:39:02.1&  0.248&  0.037& 138.4& 110.5&  0.610&  4.11&  567 $\pm$ \phn 203&  420 $\pm$  170\\
RCS~J132447+3052.2&13:24:47.20&+30:52:14.9&  0.265&  0.037&  36.1&  38.5&  0.391&  3.37&  443 $\pm$ \phn 194&  407 $\pm$  169\\
RCS~J132706+2846.8&13:27:06.10&+28:46:49.7&  0.275&  0.038&  64.8&  37.3&  0.036&  3.73&  498 $\pm$ \phn 201&  339 $\pm$  161\\
RCS~J132802+3046.1&13:28:02.40&+30:46:10.5&  0.288&  0.037&  95.1&  64.6&  0.161&  3.69&  645 $\pm$ \phn 217&  579 $\pm$  188\\
RCS~J132630+2837.8&13:26:30.80&+28:37:51.5&  0.289&  0.038&   4.6&  76.7&  0.654&  3.48&  525 $\pm$ \phn 206&  420 $\pm$  170\\
\enddata 
\end{deluxetable} 

\begin{deluxetable}{ccrlc} 
\tablecolumns{5} 
\tablecaption{Spectroscopically Confirmed Clusters in the Patch RCS1327+29}
\tablehead{ 
\colhead{Cluster}    & \colhead{RCS Redshift}    & \colhead{$\sigma_{peak}$}    &  
\colhead{Other Name}    & \colhead{Redshift}
}
\startdata 
RCS~J132532+2938.3&0.346&5.91&GHO 1323+2953&0.358\tablenotemark{1}\\
RCS~J132335+3022.6&0.508&6.19&PDCS 062     &0.467\tablenotemark{2}\\
RCS~J132421+3012.6&0.708&4.92&GHO 1322+3028&0.697\tablenotemark{2}\\
RCS~J132449+3011.6&0.755&3.66&GHO 1322+3027&0.751\tablenotemark{2}\\
\enddata 
\tablenotetext{1}{\cite{gun86}}
\tablenotetext{2}{\cite{hol99}}
\end{deluxetable}

\clearpage

See paper at {bf www.ociw.edu\/$\sim$gladders\/RCS\/papers\/survey1\/} for all 100+ Figures.

\end{document}